\documentclass[draft,showpacs,showkeys,eqsecnum,nofootinbib,aps]{revtex4}
\renewcommand{\theequation}{\arabic{equation}}
\def\beq{\begin{equation}}
\def\eeq{\end{equation}}
\def\bea{\begin{eqnarray}}
\def\eea{\end{eqnarray}}
\def\nn{\nonumber}

\begin{document}
\title{Explicit construction of BRST charge of noncommutative D-brane system}
\author{Soon-Tae Hong}
\email{soonhong@ewha.ac.kr} \affiliation{Department of Science
Education and Basic Science Research Institute,
Ewha Womans University, Seoul 120-750 Korea}
%\date{October 17, 2003}
\date{\today}%
\begin{abstract}
In the BRST-BFV scheme for noncommutative D-branes with constant
NS $B$-field, introducing ghost degrees of freedom we construct
the gauge fixed Hamiltonian and corresponding effective Lagrangian
invariant under nilpotent BRST charge. It is also shown that the
presence of auxiliary variables introduced via the improved Dirac
formalism plays a crucial role in the construction of the BRST
invariant Lagrangian.
\end{abstract}
\pacs{04.60.Ds, 11.10.Ef, 11.25.Sq, 11.30.-j} \keywords{BRST
charge, D-branes, Dirac quantization, Non-commutative geometry}
\maketitle

%%%%%%%%%%%%%%%%%%%%%%%%%%%%%%%%%%%%%%%%%%%%%%%%%%%%%%%%%%%%%%%%%%%%%%%%
\section{Introduction}
\setcounter{equation}{0}
\renewcommand{\theequation}{\arabic{section}.\arabic{equation}}
%%%%%%%%%%%%%%%%%%%%%%%%%%%%%%%%%%%%%%%%%%%%%%%%%%%%%%%%%%%%%%%%%%%%%%%%

In the matrix model compactified
on a torus, the three-form tensor field background has been shown
to be incorporated in the supersymmetric Yang-Mills theory by
deforming the base space into quantum space~\cite{connes}. This
implies that the D-brane field theory resides on a noncommutative
space in the presence of a Neveu-Schwarz (NS) $B$-field
background~\cite{douglas} due to the correspondence between the
discrete light cone quantized M-theory and the D-brane
world-volume field theory~\cite{seiberg}.  Relevance of the
noncommutative geometry in the light cone quantization of open
strings attached to D-brane has been discussed in~\cite{bigatti}.
In fact, in the matrix model the noncommutativity of spacetime is
due to non-abelian Yang-Mills gauge theory induced from parallel
$N$ D-branes at coinciding positions, all on top of each other.
The noncommutativity of spacetime takes place in the direction
perpendicular to D-branes. On the other hand, when the end points
of the open strings are attached to D-branes in the presence of a
constant NS $B$-field, the spacetime coordinates on D-branes also
do not commute in the directions parallel to D-branes. In open
string theory the noncommutativity of geometry is due to the mixed
boundary conditions, which are neither Neumann conditions nor
Dirichlet ones.  Since these boundary conditions form second-class
constraints, spacetime coordinates become noncommutative.  In a
system with noncommutative geometry, ordinary product between
functions should be replaced by the Moyal bracket~\cite{sw}.

The quantization of constraint systems has been extensively
discussed in~\cite{dirac64,hong02pr}. In particular, the embedding
of a second-class system into a first-class one~\cite{bft}, where
constraints are in strong involution, has been of much interest,
and has found a large number of applications~\cite{hong02pr}.
Recently, in the framework of Hamiltonian analysis for the
constraint systems, the noncommutativity of geometry has been
studied in~\cite{ko,hong00prd}.  The noncommutative D-brane system with
constant $B$ field has been also studied in the pp-wave
background~\cite{chu}.  Moreover, noncommutative theories has been
shown to be conventional theories embedded in a gravitational
background produced by gauge field with a charge dependent
gravitational coupling~\cite{rivelles}.  Quite recently, the
noncommutative theories have been applied to varieties of models
such as noncommutative Maxwell-Chern-Simons model~\cite{dayi} and
noncommutative Yang-Mills-Chern-Simons model~\cite{can}.

In the noncommutative D-brane systems with constant $B$ field, the
first-class Hamiltonian has been constructed to investigate its
Dirac algebra~\cite{hong00prd}. In this paper, we will construct
the first-class physical variables for this D-brane systems with
the noncommutativities, and then we will introduce canonical sets
of ghosts and anti-ghosts together with auxiliary fields to
explicitly construct the Becchi-Rouet-Stora-Tyutin (BRST) charge,
under which gauge fixed effective Lagrangian is invariant in the
BRST-Batalin-Fradkin-Vilkovisky (BFV) formalism~\cite{brst,bfv}.

%%%%%%%%%%%%%%%%%%%%%%%%%%%%%%%%%%%%%%%%%%%%%%%%%%%%%%%%%%%%%%%%%%%%%%%%
\section{Noncommutativities in D-branes with constant NS $B$-field}
\setcounter{equation}{0}
\renewcommand{\theequation}{\arabic{section}.\arabic{equation}}
%%%%%%%%%%%%%%%%%%%%%%%%%%%%%%%%%%%%%%%%%%%%%%%%%%%%%%%%%%%%%%%%%%%%%%%%

We consider the open strings whose end points are attached at
$Dp$-branes in the presence of a constant NS $B$-field. The
worldsheet action of the open string of interest, to which
background gauge fields are coupled, is given by \beq
\label{eq:open-S} S=\frac{1}{4\pi\alpha'} \int_\Sigma d^2 \sigma
\left( g_{ij}\partial_a x^i \partial^a x^j + 2\pi\alpha'
B_{ij}\epsilon^{ab}\partial_a x^i \partial_b x^j \right) + \int
d\tau \left(A_{i}\partial_{\tau}x^{i}|_{\sigma =\pi}-A_{i}
\partial_{\tau}x^{i}|_{\sigma =0} \right), \eeq
where $\Sigma$ is the string worldsheet with the metric $\eta_{ab}
= {\rm diag}~(+, -)$ and the metric of target space on D-brane is
taken to be Euclidean $g_{ij} = \delta_{ij}$ $(i,j=1,2)$.  Note
that the action (\ref{eq:open-S}) has two $U(1)$ gauge symmetries.
One of them is $\lambda$-symmetry for the transformation of $A
\rightarrow A + d\lambda$ and the other is $\Lambda$-symmetry for
the transformation with a form of both $B \rightarrow B +
d\Lambda$ and $A \rightarrow A + \Lambda$.  From the open string
action (\ref{eq:open-S}) one can readily find the equations of
motion to determine the mixed boundary conditions
\begin{equation}
  \label{eq:open-bc}
  \left. g_{ij} \partial_\sigma x^j + 2\pi\alpha' {\cal F}_{ij}
  \partial_\tau x^j \right|_{\sigma=0,\pi} = 0,
\end{equation}
where ${\cal F}=B-F$ and $F = dA$. Without background fields $B$
and $A$, the boundary conditions (\ref{eq:open-bc}) are Neumann
boundary ones, $\partial_\sigma x^i = 0$ at $\sigma =
0,\pi$, while for $B_{ij} \rightarrow \infty$ or $g_{ij}
\rightarrow 0$, the boundary conditions become Dirichlet, $\partial_\tau
x^i(\tau)=0$ on D-branes.

In order to study the boundary and the bulk near boundaries in
detail, we discretize the open string action and the boundary
conditions along the direction of $\sigma$ with equal spacing
$\epsilon = \pi/N$ which is taken to be very small with integer
$N$, and the integral for $\sigma$ being changed to the sum,
$\int_0^\pi d\sigma\rightarrow\sum_{a=0}^{N} \epsilon$.  Defining
$x_{(a)}^i$ by $x_{(a)}^i (\tau) = x^i (\tau, \sigma)|_{\sigma=a
\epsilon} $, we have the discretized Lagrangian of the open string
action (\ref{eq:open-S}) as follows \bea L_{0}
&=&\frac{1}{4\pi\alpha'}\sum_{a=0} \left(\epsilon
\left(\dot{x}_{(a)}^i\right)^2 - \frac{1}{\epsilon}
   \left( x_{(a+1)}^i - x_{(a)}^i \right)^2 + 4\pi\alpha' B_{ij}
   \dot{x}_{(a)}^i \left( x_{(a+1)}^j - x_{(a)}^j \right)\right)\nn\\
   & &+\sum_{a=0} A_i (\dot{x}_{(a+1)}^i -\dot{x}_{(a)}^i),
\label{eq:d-S} \eea where the overdot denotes derivative with
respect to $\tau$ and the mixed boundary conditions are given by
\begin{equation}
  \label{eq:d-bc}
  g_{ij} \frac{1}{\epsilon} (x_{(1)}^j - x_{(0)}^j)
  + 2\pi\alpha' {\cal F}_{ij} \dot{x}_{(0)}^j = 0,
\end{equation}
with $x_{(0)}^i$ denoting the end of the open strings.  Here we
have taken only $\sigma=0$ case of the boundary conditions since
we are interested in the boundary $\sigma = 0$ only.  Note that
the action (\ref{eq:d-S}) and the boundary conditions
(\ref{eq:d-bc}) become continuous in $\sigma$-direction in the
limit that $N$ goes to infinity, or equivalently $\epsilon$ is
very small.

From the action (\ref{eq:d-S}) we obtain the canonical momenta of
$x_{(a)}^i$ $(a = 0, 1,2, \cdots)$
\begin{equation}
\label{eq:d-mom}
p_{(a)i} = \frac{\epsilon}{2\pi\alpha'}
  \left( g_{ij}\dot{x}_{(a)}^j +\frac{2\pi\alpha'}{\epsilon}B_{ij}
  \left(x_{(a+1)}^j - x_{(a)}^j \right)
-\frac{2\pi\alpha'}{\epsilon}A_i \delta_{a0} \right).
\end{equation}
The combination of the boundary conditions (\ref{eq:d-bc}) and the
canonical momenta (\ref{eq:d-mom}) gives the primary
constraints~\cite{hong00prd}
\begin{equation}
\label{eq:d-constr}
\Omega_i = \frac{1}{\epsilon}\left((2\pi\alpha')^2 {\cal F}_{ij}
(p_{(0)}^j + A^j) + M_{ij} (x_{(1)}^j - x_{(0)}^j )\right) \approx 0,
\end{equation}
where $M_{ij} = \left[g- (2\pi\alpha')^2 {\cal
F}g^{-1}B\right]_{ij}$. By taking the Legendre transformation of
the Lagrangian (\ref{eq:d-S}) we can obtain the total Hamiltonian
of the form $H_T = H+u^i \Omega_i$ where the canonical
Hamiltonian $H$ and the Lagrangian multipliers $u^i$ are given by
\bea H&=&\frac{1}{4\pi\alpha'\epsilon}\sum_{a=1}
\left((2\pi\alpha')^2 \left(p_{(a)i}- B_{ij} (x_{(a+1)}^j -
x_{(a)}^j)\right)^2 + (x_{(a+1)}^i - x_{(a)}^i)^2 \right)\label{eq:d-Hc}\\
&&+\frac{1}{4\pi\alpha'\epsilon}[\theta^{-1}{\cal F}^{-1}g]_{ij}
(x_{(1)}^i - x_{(0)}^i)(x_{(1)}^j - x_{(0)}^j),\nn\\
u^i &=& \frac{2\pi\alpha'}{\epsilon^2 \Delta_{12}}\epsilon^{ij}
\left(M_{jk} p_{(1)}^k - [Mg^{-1}B]_{jk} (x_{(2)}^k - x_{(1)}^k) +
\theta_{jk}^{-1} (x_{(1)}^k - x_{(0)}^k) \right), \label{ui} \eea
with $\theta_{ij}^{-1} = - \frac{1}{(2\pi\alpha')^2} \left[ (g -
  2\pi\alpha' {\cal F}) {\cal F}^{-1} (g + 2\pi\alpha' {\cal F})
  \right]_{ij}$.  Note that with the Lagrangian multipliers (\ref{ui}) one can
obtain the strongly involutive constraint algebra $\{\Omega_i,
H_{T}\}=0$, and the constraints (\ref{eq:d-constr}) form a
second-class system since their Poisson brackets are given by
\beq
\label{eq:d-PB}
\Delta_{ij}=\{ \Omega_i, \Omega_j \}
=\frac{(2\pi\alpha')^2}{\epsilon^2} \left[(G + M)g^{-1}{\cal
F}\right]_{ij}, \eeq
where $G_{ij} = \left[ (g + 2\pi\alpha' {\cal F}) g^{-1} (g - 2\pi\alpha'
  {\cal F}) \right]_{ij}$.

%%%%%%%%%%%%%%%%%%%%%%%%%%%%%%%%%%%%%%%%%%%%%%%%%%%%%%%%%%%%%%%%%%%%%%%%
\section{First-class physical variables in D-branes with constant NS $B$-field}
\setcounter{equation}{0}
\renewcommand{\theequation}{\arabic{section}.\arabic{equation}}
%%%%%%%%%%%%%%%%%%%%%%%%%%%%%%%%%%%%%%%%%%%%%%%%%%%%%%%%%%%%%%%%%%%%%%%%

Next, in order to construct the BRST invariant effective Lagrangian, we need to
find the first-class physical variables, which enables us to apply the
standard BRST-BFV formalism~\cite{brst,bfv} to the constrained systems
as in unconstrained systems.  Recall that the D-brane system of interest has
the second-class constraints in the boundary conditions, which yields
the noncommutativity of spacetime on the D-brane.

Following the Batalin-Fradkin-Tyutin scheme~\cite{bft,hong02pr},
we systematically convert the second-class constraints algebra
involved in the noncommutative D-brane system into a strongly
involutive one by introducing a pair of canonically conjugate
auxiliary variables $y_i$ ($i= 1,2$) with the Poisson brackets
\beq \{y_{i}, y_{j}\} = \epsilon_{ij}.\eeq In this enlarged phase
space one systematically constructs the first-class constraints
$\tilde{\Omega}_{i}$ as a power series in these auxiliary
variables, by requiring that they be in strong involution
$\{\tilde{\Omega}_{i},\tilde{\Omega}_{j}\}=0$~\cite{hong00prd},
\beq \tilde{\Omega}_{1}=\Omega_{1}+y_{1},~~~
\tilde{\Omega}_{2}=\Omega_{2}-\Delta_{12} y_{2}.
\label{eq:m-constr} \eeq

In this extended phase we construct the first-class physical
variables $\tilde{{\cal F}} =(\tilde{q}_{a},\tilde{p}_{a})$,
corresponding to the original physical  variables defined by
${\cal F}=(q_{a},p_{a})$ . They are again obtained as a power
series in the auxiliary fields $(y_{i}, y_{j})$ by demanding that
they be in strong involution with the first-class constraints
(\ref{eq:m-constr}), that is $\{\tilde{\Omega}_{i}, \tilde{{\cal
F}}\}=0$.  After some tedious algebra, we obtain for the
first-class physical variables\footnote{In Ref.~\cite{hong00prd},
the first-class variables (\ref{pitilde}) and their corresponding
Dirac algebra (\ref{tconsts}) and (\ref{commstd}) have not been
explicitly constructed.  Moreover, the ghosts and anti-ghosts, and
their corresponding BRST charges, which will be discussed in the
next section, have not been introduced in Ref.~\cite{hong00prd}.}
\bea
\tilde{x}^{i}_{(a)}&=&x^{i}_{(a)}+\frac{(2\pi\alpha^{\prime})^{2}}{\epsilon}
{\cal F}_{12}\delta_{a0}\left(g^{i2}y_{2}-g^{i1}\frac{y_{1}}{\Delta_{12}}\right),\nn\\
\tilde{p}_{(a)i}&=&p_{(a)i}+\frac{1}{2\epsilon}\left((G_{11}+M_{11})\delta_{a0}
-2M_{11}\delta_{a1}\right)\left(g_{i1}y_{2}+g_{i2}\frac{y_{1}}{\Delta_{12}}\right).
\label{pitilde}
\end{eqnarray}
In terms of these physical variables the first-class Hamiltonian can be
written in the compact form
\bea
\tilde{H}&=&\frac{1}{4\pi\alpha'\epsilon}\sum_{a=1}
\left((2\pi\alpha')^2 \left(\tilde{p}_{(a)i}- B_{ij}
(\tilde{x}_{(a+1)}^j -\tilde{x}_{(a)}^j)\right)^2
+(\tilde{x}_{(a+1)}^i - \tilde{x}_{(a)}^i)^2 \right)\nn\\
& &+\frac{1}{4\pi\alpha'\epsilon}[\theta^{-1}{\cal F}^{-1}g]_{ij}
(\tilde{x}_{(1)}^i - \tilde{x}_{(0)}^i)(\tilde{x}_{(1)}^j
-\tilde{x}_{(0)}^j). \label{htilde} \eea We then directly rewrite
this Hamiltonian in terms of the original and auxiliary variables
\bea
\tilde{H}&=&H-\frac{2\pi\alpha'}{\epsilon^2}\frac{y_{1}}{\Delta_{12}}
\left(M_{22} p_{(1)}^2 - [Mg^{-1}B]_{21} (x_{(2)}^1 - x_{(1)}^1) +
\theta_{21}^{-1} (x_{(1)}^1 - x_{(0)}^1) \right)\nn\\
& & \quad\ \, - \frac{2\pi\alpha'}{\epsilon^2}y_{2} \left(M_{11}
p_{(1)}^1 -[Mg^{-1}B]_{12} (x_{(2)}^2 - x_{(1)}^2) +
\theta_{12}^{-1}(x_{(1)}^2 - x_{(0)}^2) \right)\nn\\
& & \quad\ \, + \frac{\pi\alpha'}{\epsilon^3} [G + Mg^{-1}M]_{11}
\left(\frac{y_{1}^{2}}{\Delta_{12}^2}+y_{2}^2\right) \label{hctp} \eea
which is strongly involutive with the first-class constraints
$\{\tilde{\Omega}_{i},\tilde{H}\}=0$.

Now, we consider the Poisson brackets of physical variables in the
extended phase space $\tilde{{\cal F}}$ and identify the Dirac
brackets by taking the vanishing limit of auxiliary variables. After
some manipulation, from (\ref {pitilde}), one readily finds the
commutators
\begin{eqnarray}
\{\tilde{x}_{(a)}^i, \tilde{x}_{(b)}^j \}&=& - (2\pi\alpha')^2
      \left[ (G+M)^{-1}
      {\cal F} g^{-1} \right]^{ij}\delta_{a0}\delta_{b0}, \nn\\
\{\tilde{x}_{(a)}^i, \tilde{p}_{(b)j} \} &=& \delta^{i}_{j}
      (\delta_{ab}-\frac{1}{2}\delta_{a0}\delta_{b0})
      +\left[(G+M)^{-1}M\right]^{i}_{j}\delta_{a0}\delta_{b1}, \nn\\
\{\tilde{p}_{(a)i}, \tilde{p}_{(b)j} \} &=&
\frac{1}{(2\pi\alpha')^{2}}
      \left(\frac{1}{4}\left[g{\cal F}^{-1}(G+M)\right]_{ij}\delta_{a0}\delta_{b0}
+\left[g{\cal F}^{-1}M(G+M)^{-1}M\right]_{ij}\delta_{a1}\delta_{b1}\right.\nn\\
  & &\left. -\frac{1}{2}(g{\cal F}^{-1}M)_{ij}(\delta_{a0}\delta_{b1}
      +\delta_{a1}\delta_{b0})\right).
\label{tconsts}
\end{eqnarray}
Note that the above Poisson brackets in the extended phase space
exactly reproduce the corresponding Dirac brackets
\begin{eqnarray}
\{\tilde{x}_{(a)}^i,\tilde{x}_{(b)}^j\}&=& \{x_{(a)}^i,x_{(b)}^j\}_{D},\nn\\
\{\tilde{x}_{(a)}^i,\tilde{p}_{(b)}^j\}&=& \{x_{(a)}^i,p_{(b)}^j\}_{D},
\nonumber\\
\{\tilde{p}_{(a)}^i,\tilde{p}_{(b)}^j\}&=& \{p_{(a)}^i,p_{(b)}^j\}_{D},
\label{commstd}
\end{eqnarray}
where $\{A,B\}_{D}=\{A,B\}-\{A,\Omega_{k}\}\Delta^{k k^{\prime}}
\{\Omega_{k^{\prime}},B\}$ with $\Delta^{k k^{\prime}}$ being the
inverse of $\Delta_{k k^{\prime}}$ in (\ref{eq:d-PB}). Also it is
amusing to see in (\ref{commstd}) that these Poisson brackets of
$\tilde{{\cal F}}$'s have exactly the same form of
the Dirac brackets of the field ${\cal F}$ obtained by the replacement of $%
{\cal F}$ with $\tilde{{\cal F}}$, namely
$\{\tilde{A},\tilde{B}\}=\{A,B\}_{D}|_{A\rightarrow
\tilde{A},B\rightarrow \tilde{B}}$.

%%%%%%%%%%%%%%%%%%%%%%%%%%%%%%%%%%%%%%%%%%%%%%%%%%%%%%%%%%%%%%%%%%%%%%%%
\section{BRST invariant effective Lagrangian}
\setcounter{equation}{0}
\renewcommand{\theequation}{\arabic{section}.\arabic{equation}}
%%%%%%%%%%%%%%%%%%%%%%%%%%%%%%%%%%%%%%%%%%%%%%%%%%%%%%%%%%%%%%%%%%%%%%%%

In this section, in order to obtain the BRST invariant effective
Lagrangian in the framework of the BRST-BFV
formalism~\cite{brst,bfv}, which is applicable to theories with
the first-class constraints discussed above, we introduce two
canonical sets of ghosts and anti-ghosts together with auxiliary
fields  $({\cal C}^{i},\bar{{\cal P}}_{i}),~({\cal P}^{i},
\bar{{\cal C}}_{i}), ~({\cal N}^{i},{\cal B}_{i})$, $(i=1,2)$ which satisfy the
super-Poisson algebra \beq \{{\cal C}^{i},\bar{{\cal
P}}_{j}\}=\{{\cal P}^{i},
  \bar{{\cal C}}_{j}\}=\{{\cal N}^{i},{\cal B}_{j}\}=\delta_{j}^{i},
\eeq with the super-Poisson bracket defined as \beq
\{A,B\}=\frac{\delta A}{\delta q}|_{r}\frac{\delta B}{\delta
p}|_{l}
-(-1)^{\eta_{A}\eta_{B}}\frac{\delta B}{\delta q}|_{r}\frac{\delta A} {%
\delta p}|_{l}. \eeq Here $\eta_{A}$ denotes the number of
fermions called ghost number in $A$ and the subscript $r$ and $l$
imply right and left derivatives, respectively.

In the noncommutative D-brane system, we now construct the BRST
charge $Q$ and the fermionic gauge fixing function $\Psi$,
together with the unitary gauge choice $\chi^{1}=\Omega_{1}$,
$\chi^{2}=\Omega_{2}$, \bea Q&=&{\cal
C}^{i}\tilde{\Omega}_{i}+{\cal
P}^{i}{\cal B}_{i},\label{brstch0}\\
\Psi&=&\bar{{\cal C}}_{i}\chi^{i}+\bar{{\cal P}}_{i}{\cal N}^{i},
\label{brstcharge} \eea with the property $Q^{2}=\{Q,Q\}=0$. This
nilpotent charge $Q$ is the generator of the following
infinitesimal transformations, \beq
\begin{array}{ll}
\delta_{Q} x_{(a)}^{i} = \frac{1}{\epsilon} (2\pi\alpha')^2
\delta_{a0}{\cal F}_{ij}{\cal C}^j, &\\
\delta_{Q} p_{(a)i} = -\frac{1}{2\epsilon}\delta_{a0}
\left(G_{ij}-M_{ij}\right){\cal C}^j
+\frac{1}{\epsilon}(\delta_{a1}-\delta_{a0})M_{ij}{\cal C}^j, &\\
\delta_{Q} y_{1} = \Delta_{12}{\cal C}^2, &\delta_{Q} y_{2} =
{\cal C}^1,\\
\delta_{Q} \bar{\cal P}_{i}= \tilde{\Omega}_{i}, &\delta_{Q} {\cal C}^{i} = 0,\\
\delta_{Q} \tilde{\cal C}_{i} = {\cal B}_{i}, &\delta_{Q} {\cal P}^{i}= 0,\\
\delta_{Q} {\cal B}_{i} = 0, &\delta_{Q} {\cal N}^i = -{\cal P}^i,\\
\end{array}
\label{eq:transf-x} \eeq under which the first-class Hamiltonian
$\tilde{H}$ and $\{Q,\Psi\}$ are invariant to yield \bea
\delta_{Q}\tilde{H}&=&\{Q,\tilde{H}\}=0,\nn\\
\delta_{Q}\{Q,\Psi\}&=&\{Q,\{Q,\Psi\}\}=0. \label{qh} \eea The
``gauge fixed" effective Hamiltonian is then given by \beq
H_{eff}=\tilde{H}-\{Q,\Psi\}=\tilde{H}-\Delta_{12}({\cal
C}^{1}\bar{\cal C}_{2}-{\cal C}^{2}\bar{\cal C}_{1})
-\tilde{\Omega}_{i}{\cal N}^{i}-\Omega_{i}{\cal B}_{i}-\bar{\cal
P}_{i}{\cal P}^{i},\label{heff} \eeq where $\tilde{H}$ is given by
(\ref{eq:d-Hc}) and (\ref{hctp}).  This Hamiltonian is invariant
under the transformation (\ref{eq:transf-x}) with the BRST charge
(\ref{brstch0}).

After some algebra associated with the Legendre transformation of $H_{eff}$,
we arrive at the effective quantum Lagrangian of the form
\begin{equation}
L_{eff}= L_{0} + L_{WZ} + L_{ghost},
\end{equation}
where $L_{0}$ is given by (\ref{eq:d-S}) and
\begin{eqnarray}
L_{WZ}&=&2\pi\alpha'{\cal F}_{12}\left(-\frac{\dot{y}_{1}}{\Delta_{12}}\dot{x}_{(0)}^{1}
+\dot{y}_{2}\dot{x}_{(0)}^{2}+\frac{(2\pi\alpha')^{2}}{2\epsilon}{\cal F}_{12}
\left(\frac{\dot{y}_{1}^{2}}{\Delta_{12}^{2}}
+\dot{y}_{2}^{2}\right)\right)\nn\\
& &+\frac{2\pi\alpha'}{\epsilon^{2}}{\cal F}_{12}
\left(-\frac{y_{1}}{\Delta_{12}}(x_{(1)}^{1}-x_{(0)}^{1})
+y_{2}(x_{(1)}^{2}-x_{(0)}^{2})-\frac{(2\pi\alpha')^{2}}{2\epsilon}{\cal F}_{12}
\left(\frac{y_{1}^{2}}{\Delta_{12}^{2}}
+y_{2}^{2}\right)\right)\nn\\
& &+\frac{(2\pi\alpha')^{2}}{\epsilon}{\cal F}_{12}B_{12}
\left(-\frac{\dot{y}_{1}}{\Delta_{12}}(x_{(1)}^{2}-x_{(0)}^{2})
-y_{2}\dot{x}_{(0)}^{1}+\frac{(2\pi\alpha')^{2}}{\epsilon}{\cal F}_{12}
\frac{\dot{y}_{1}}{\Delta_{12}}y_{2}\right)\nn\\
& &+\frac{(2\pi\alpha')^{2}}{\epsilon}{\cal F}_{12}B_{21}
\left(\frac{y_{1}}{\Delta_{12}}\dot{x}_{(0)}^{2}+\dot{y}_{2}(x_{(1)}^{1}-x_{(0)}^{1})
+\frac{(2\pi\alpha')^{2}}{\epsilon}{\cal F}_{12}
\frac{y_{1}}{\Delta_{12}}\dot{y}_{2}\right)\nn\\
& &+\frac{(2\pi\alpha')^{2}}{\epsilon}{\cal F}_{12}
\left(\frac{\dot{y}_{1}}{\Delta_{12}}A_{1}-\dot{y}_{2}A_{2}\right),\nn\\
L_{ghost}&=&-\frac{1}{\Delta_{12}}\dot{\cal
B}_{2}\dot{y}_{1}+\dot{\bar{\cal C}}_{2}\dot{\cal C}^{2}.
\label{lafgh}
\end{eqnarray}
This Lagrangian is invariant under the BRST transformation
\beq
\begin{array}{ll}
\delta_{\lambda} x_{(a)}^{i}=
-\lambda\frac{(2\pi\alpha')^2}{\epsilon}\delta_{a0} {\cal
F}_{ij}{\cal C}^j, &\delta_{\lambda}
y_{1}=-\lambda\Delta_{12}{\cal C}^2,\\
\delta_{\lambda} y_{2} =-\lambda{\cal C}^1, &\delta_{\lambda}
\bar{\cal C}_{i}=-\lambda {\cal B}_{i},\\
\delta_{\lambda} {\cal C}^{i}=0, &\delta_{\lambda} {\cal B}_{i}=0,
\end{array}
\eeq
where $\lambda$ is an infinitesimal Grassmann valued
parameter.

%%%%%%%%%%%%%%%%%%%%%%%%%%%%%%%%%%%%%%%%%%%%%%%%%%%%%%%%%%%%%%%%%%%%%%%%
\section{Conclusions}
\setcounter{equation}{0}
\renewcommand{\theequation}{\arabic{section}.\arabic{equation}}
%%%%%%%%%%%%%%%%%%%%%%%%%%%%%%%%%%%%%%%%%%%%%%%%%%%%%%%%%%%%%%%%%%%%%%%%

In conclusion, we have constructed the first-class Hamiltonian for
the D-brane systems with constant NS $B$-field in the
Batalin-Fradkin-Tyutin scheme. In the BRST-BFV formalism, we have
next introduced canonical sets of ghosts and anti-ghosts together
with auxiliary fields and thus we have constructed their BRST
invariant first-class Hamiltonian and the corresponding effective
Lagrangian which is invariant under transformation with nilpotent
BRST charge.  Moreover it has been shown that the presence of
auxiliary variables introduced via the improved Dirac formalism
plays a crucial role in the construction of the BRST invariant
Lagrangian, since the BRST-BFV formalism is applicable to theories
with the first-class constraints associated with these auxiliary
variables.

It would be desirable if the continuum limit of our discretized
BRST invariant effective Lagrangian can be formulated.  This work
is in progress and will be reported elsewhere.

\acknowledgments The author would like to thank the referee for helpful comments.
This work was supported by the Korea Science and Engineering Foundation Grant R01-2000-00015.

%%%%%%%%%%%%%%%%%%%%%%%%%% REFERENCES %%%%%%%%%%%%%%%%%%%%%%%%%%%%%%%%


\begin{thebibliography}{99}
\bibitem{connes} A. Connes, M.R. Douglas, A. Schwarz, JHEP 02 (1998) 003.
\bibitem{douglas} M.R. Douglas, C. Hull, JHEP 02 (1998) 008.
\bibitem{seiberg}  N. Seiberg, Phys. Rev. Lett. 79 (1997) 3577.
\bibitem{bigatti} D. Bigatti, L. Susskind, Phys. Rev. D 62 (2000) 066004.
\bibitem{sw} N. Seiberg, E. Witten, JHEP 09 (1999) 032.
\bibitem{dirac64} P.A.M. Dirac, {\it Lectures on Quantum Mechanics} (Yeshiba University Press,
New York, 1964).
\bibitem{hong02pr} S.T. Hong, Y.J. Park, Phys. Rep. 358 (2002) 143, and references therein.
\bibitem{bft} I.A. Batalin, E.S. Fradkin, Nucl. Phys. B 279 (1987) 514;
          I.A. Batalin, I.V. Tyutin, Int. J. Mod. Phys. A 6 (1991) 3255.
\bibitem{ko} C.S. Chu, P.M. Ho, Nucl. Phys. B 550 (1999) 151;
          T. Lee, Phys. Rev. D 62 (2000)
          024022; C.S. Chu, P.M. Ho, Nucl. Phys. B 568 (2000) 447.
\bibitem{hong00prd} S.T. Hong, W.T. Kim, Y.J. Park, M.S. Yoon, Phys. Rev. D 62 (2000) 085010.
\bibitem{chu} C.S. Chu, P.M. Ho, Nucl. Phys. B 636 (2002) 141.
\bibitem{rivelles} V.O. Rivelles, Phys. Lett. B 558 (2003) 191.
\bibitem{dayi} O.F. Dayi, Phys. Lett. B 560 (2003) 239.
\bibitem{can} M.B. Cantcheff, P. Minces, Phys. Lett. B 557 (2003) 283.
\bibitem{brst} C. Becchi, A. Rouet, R. Stora, Ann. Phys. 98 (1976) 287;
I.V. Tyutin, Lebedev Preprint 39 (1975) unpublished.
\bibitem{bfv} E.S. Fradkin, G.A. Vilkovisky, Phys. Lett. B 55 (1975) 224;
          M. Henneaux, Phys. Rep. 126 (1985) 1.
\end{thebibliography}
\end{document}